\documentclass{article}
\usepackage[T1]{fontenc} 
\usepackage[utf8]{inputenc} 
\usepackage{ismir,amsmath,cite,url}
\usepackage{graphicx}
\usepackage{color}
\usepackage{dsfont}
\usepackage[nolist]{acronym}
\usepackage[capitalise, noabbrev]{cleveref}
\crefname{equation}{}{}

\usepackage{mathtools}

\usepackage{siunitx}
\usepackage{caption}
\usepackage{subcaption}
\usepackage{pgfplots}
\usepgfplotslibrary{groupplots,units}
\usepackage[export]{adjustbox}

\usepackage{float}
\usepackage{placeins}
\usepackage{todonotes}

\usepackage{lineno}

\title{Stabilizing Training with Soft Dynamic Time Warping:\\A Case Study for Pitch Class Estimation with Weakly Aligned Targets}

\multauthor{Johannes Zeitler \hspace{1cm} Simon Deniffel \hspace{1cm} Michael Krause \hspace{1cm} Meinard Müller}{International Audio Laboratories Erlangen, Germany\\\small\tt \{johannes.zeitler, michael.krause, meinard.mueller\}@audiolabs-erlangen.de}

\def\authorname{J. Zeitler, S. Deniffel, M. Krause, and M. Müller}

\begin{document}

\begin{acronym}
\acro{CTC}[CTC]{connectionist temporal classification}
\acro{MCTC}[MCTC]{multi-label connectionist temporal classification}
\acro{SAE}[SAE]{soft alignment error}
\acro{SDTW}[SDTW]{soft dynamic time warping}
\acro{DTW}[DTW]{dynamic time warping}
\acro{DNN}[DNN]{deep neural network}
\acro{MSE}[MSE]{mean squared error}
\acro{MIR}[MIR]{music information retrieval}
\acro{PCE}[PCE]{pitch class estimation}
\acro{MPE}[MPE]{multi-pitch estimation}
\acro{HCQT}[HCQT]{harmonic constant-Q transform}
\acro{CNN}[CNN]{convolutional neural network}
\acro{SWD}[SWD]{Schubert Winterreise dataset}
\end{acronym}

\maketitle
\begin{abstract}
\Ac{SDTW} is a differentiable loss function that allows for training neural networks from weakly aligned data. Typically, \ac{SDTW} is used to iteratively compute and refine soft alignments that compensate for temporal deviations between the training data and its weakly annotated targets. 
One major problem is that a mismatch between the estimated soft alignments and the reference alignments in the early training stage leads to incorrect parameter updates, making the overall training procedure unstable.
In this paper, we investigate such stability issues by considering the task of pitch class estimation from music recordings as an illustrative case study. In particular, we introduce and discuss three conceptually different strategies (a hyperparameter scheduling, a diagonal prior, and a sequence unfolding strategy) with the objective of stabilizing intermediate soft alignment results.
Finally, we report on experiments that demonstrate the effectiveness of the strategies and discuss efficiency and implementation issues.
\end{abstract}

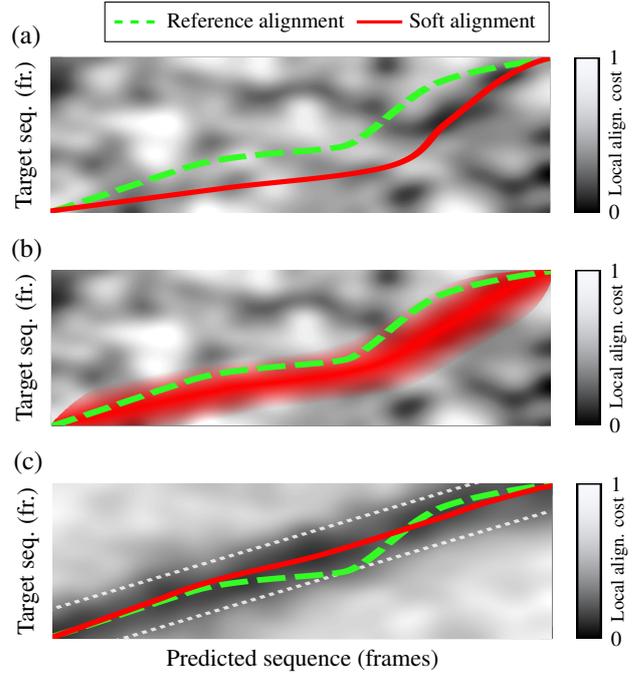
\begin{figure}[t!]
     \centering
     \hspace{-.3cm}
\begin{subfigure}[b]{\columnwidth}
        \centering
        \begin{tikzpicture}
        	\begin{axis}[
        		width=.8\columnwidth,
        		height=.25\columnwidth,
        		scale only axis,
        		enlargelimits=false,
        		xtick={},
        		xmin=0,
        		xmax=1,
        		xticklabels={},
                yticklabels={},
                ytick={},
        		xlabel={},
        		ylabel={Target seq. (fr.)},
        		y label style={yshift=-.9cm, font=\small},
        		clip=false,
                legend style={at={(1.0,1.1)}, anchor=south east, font=\footnotesize, legend columns=2},
                legend cell align=left,
        		]

                \addlegendimage{dashed, line legend, green, style={very thick}};
                \addlegendimage{line legend, red, style={very thick}};
          
        		\addplot graphics[xmin=0,xmax=1,ymin=0,ymax=2] {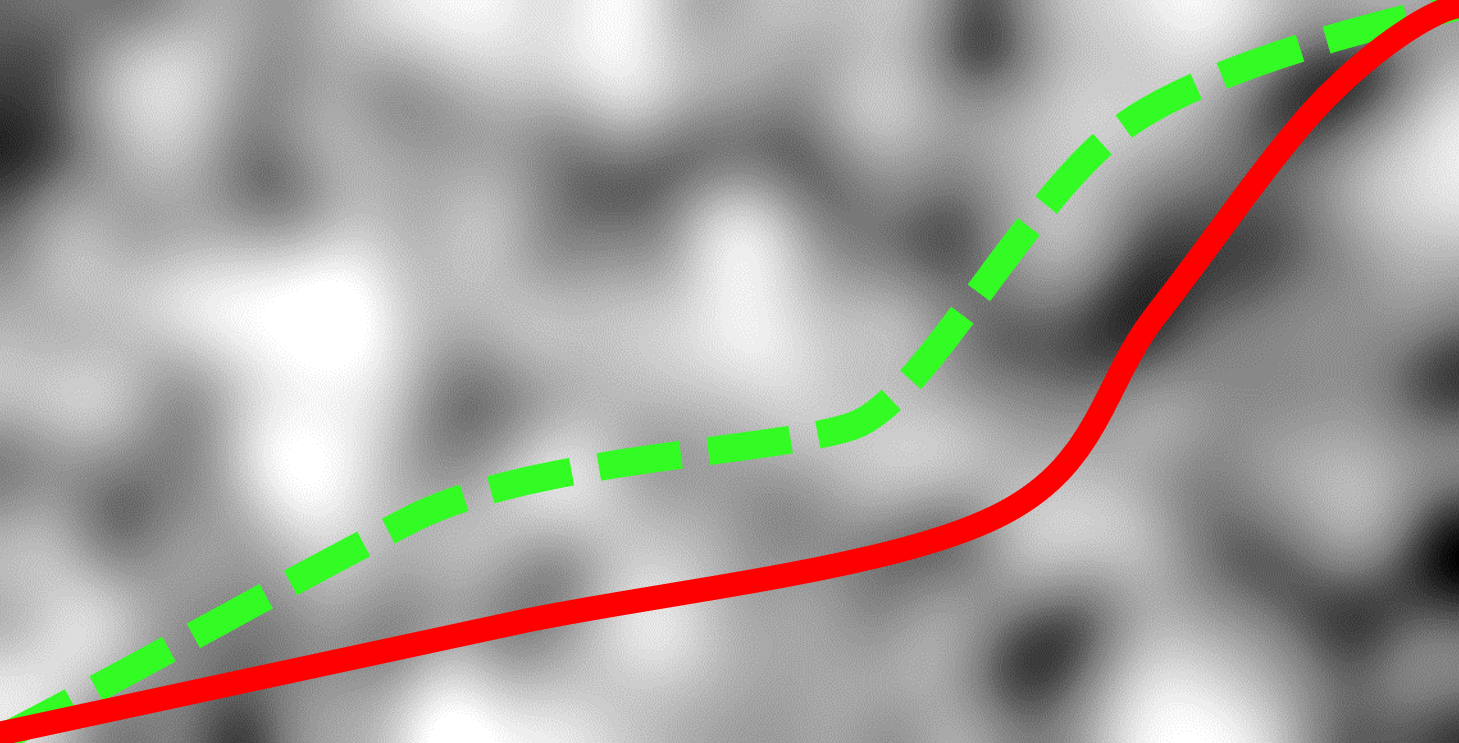};
        		\addplot graphics[xmin=1.05,xmax=1.1,ymin=0,ymax=2] {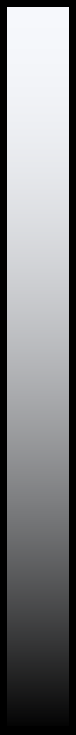};
        		\node[anchor=west, font=\footnotesize] at (axis cs:1.1,0) {0};
        		\node[anchor=west, font=\footnotesize] at (axis cs:1.1,2) {1};
          
          \node[anchor=north, rotate=90, align=center, font=\scriptsize] at (axis cs:1.1,1.02) {Local align. cost};

          \node[anchor=south, align=right] at (axis cs: -.05, 2.0) {(a)};

            \legend{Reference alignment~~,
                    Soft alignment};

        	\end{axis}
        \end{tikzpicture}
        \label{fig:teaser_mismatch}
     \end{subfigure}
     \vfill
     \vspace{-1cm}
     \hspace{-.3cm}
     \begin{subfigure}[b]{\columnwidth}
         \centering
         \begin{tikzpicture}
        	\begin{axis}[
        		width=.8\columnwidth,
        		height=.25\columnwidth,
        		scale only axis,
        		enlargelimits=false,
        		xtick={},
        		xmin=0,
        		xmax=1,
        		xticklabels={},
                yticklabels={},
                ytick={},
        		xlabel={},
        		ylabel={Target seq. (fr.)},
        		y label style={yshift=-.9cm, font=\small},
        		clip=false
        		]
        		\addplot graphics[xmin=0,xmax=1,ymin=0,ymax=2] {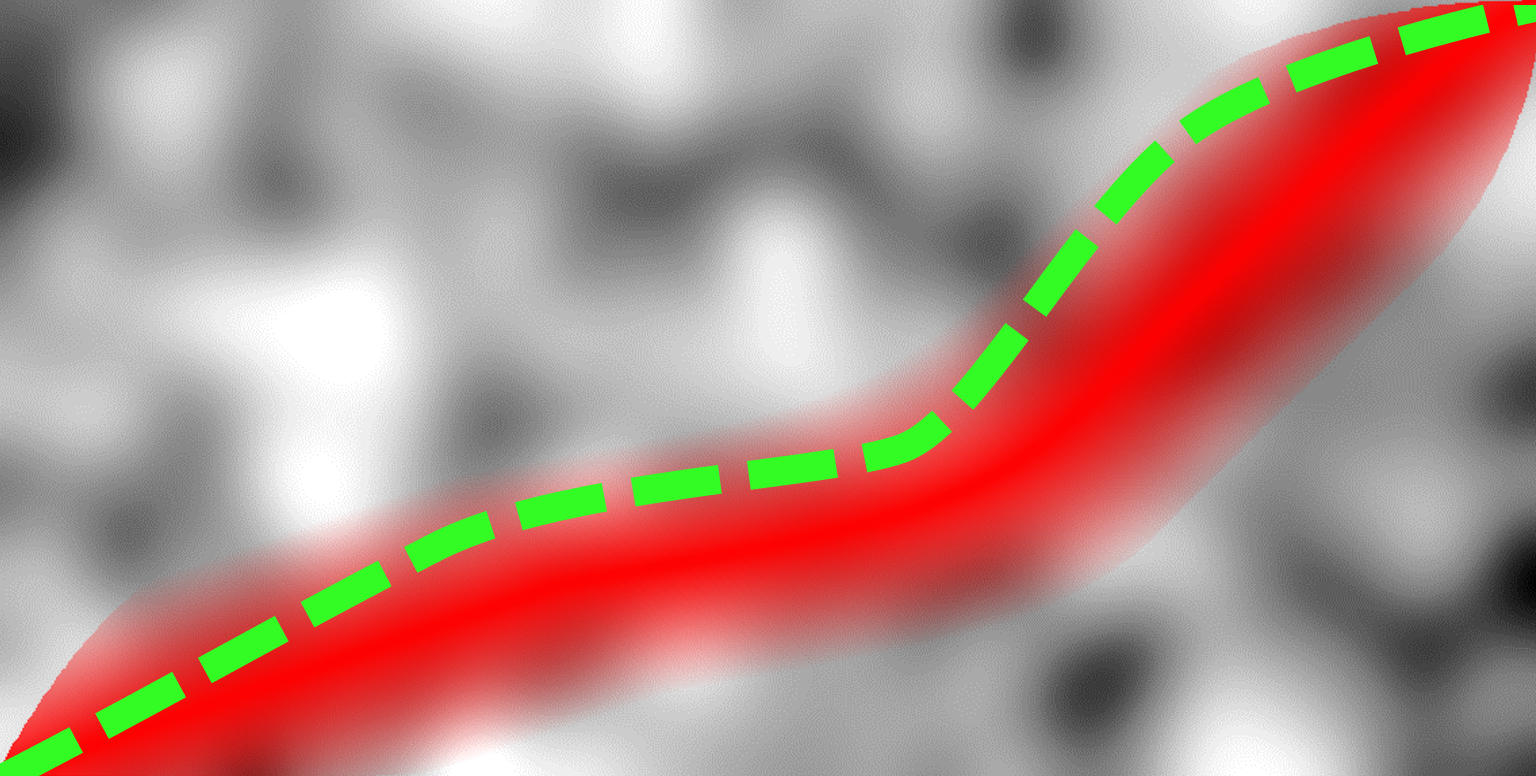};
        		\addplot graphics[xmin=1.05,xmax=1.1,ymin=0,ymax=2] {figs/cbar_bw_inv};
        		\node[anchor=west, font=\footnotesize] at (axis cs:1.1,0) {0};
        		\node[anchor=west, font=\footnotesize] at (axis cs:1.1,2) {1};
          \node[anchor=north, rotate=90, align=center, font=\scriptsize] at (axis cs:1.1,1.02) {Local align. cost};
          \node[anchor=south, align=right] at (axis cs: -.05, 2.0) {(b)};
        	\end{axis}
        \end{tikzpicture}
         \label{fig:teaser_smooth}
     \end{subfigure}
     \vfill
     \vspace{-1cm}
     \hspace{-.241cm}
     \begin{subfigure}[b]{\columnwidth}
         \centering
         \begin{tikzpicture}
        	\begin{axis}[
        		width=.8\columnwidth,
        		height=.25\columnwidth,
        		scale only axis,
        		enlargelimits=false,
        		xtick={},
        		xmin=0,
        		xmax=1,
        		xticklabels={},
                yticklabels={},
                ytick={},
        		xlabel={Predicted sequence (frames)},
        		ylabel={Target seq. (fr.)},
                x label style={yshift=.5cm, font=\small},
        		y label style={yshift=-.9cm, font=\small},
        		clip=false
        		]
        		\addplot graphics[xmin=0,xmax=1,ymin=0,ymax=2] {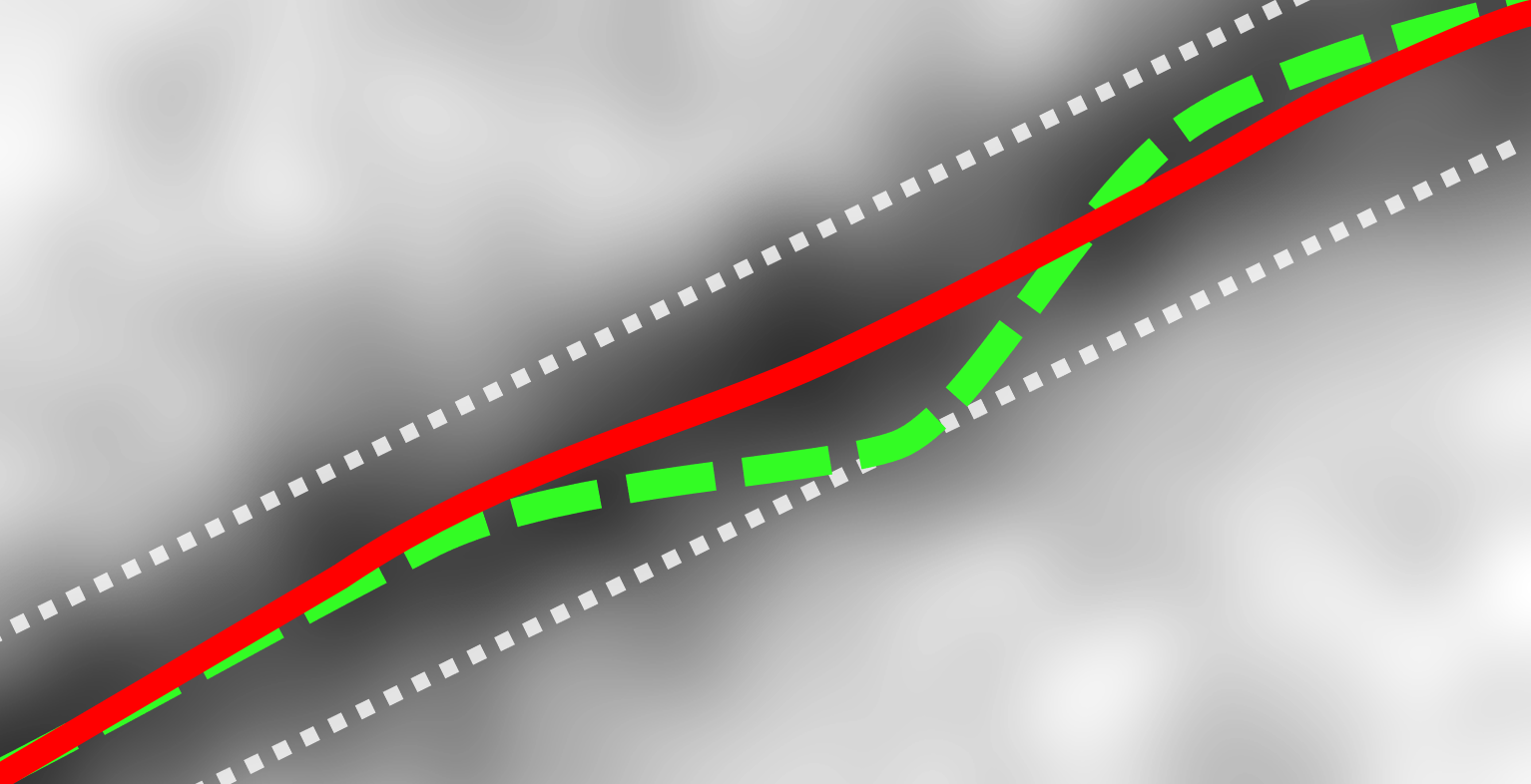};
        		\addplot graphics[xmin=1.05,xmax=1.1,ymin=0,ymax=2] {figs/cbar_bw_inv};
        		\node[anchor=west, font=\footnotesize] at (axis cs:1.1,0) {0};
        		\node[anchor=west, font=\footnotesize] at (axis cs:1.1,2) {1};
          \node[anchor=north, rotate=90, align=center, font=\scriptsize] at (axis cs:1.1,1.02) {Local align. cost};
          \node[anchor=south, align=right] at (axis cs: -.05, 2.0) {(c)};
        	\end{axis}
        \end{tikzpicture}
         \label{fig:teaser_diag}
     \end{subfigure}
     \vspace{-0.7cm}
        \caption{Deviation of strong reference alignments (dashed green) and soft alignments (red) and stabilizing strategies. \textbf{(a)}~Alignment mismatch of standard \acs{SDTW}. Stabilizing alignments with \textbf{(b)}~hyperparameter scheduling and \textbf{(c)}~diagonal prior.}
        \label{fig:teaser}
        \vspace{-.3cm}
\end{figure}

\acresetall
\section{Introduction and Related Work}
\Acp{DNN} have been commonly used in many \ac{MIR} tasks, such as music transcription~\cite{BenetosDDE19_MusicTranscription_SPM}, or \ac{PCE}~\cite{KorezeniowskiW16_DeepChroma_ISMIR,WeissZZSM21_DeepChromaChord_ISMIR}. The latter provides a widely-used feature representation for various subsequent processing pipelines, e.g., audio thumbnailing~\cite{BartschW05_chroma_IEEEMULTIMEDIA}, or chord recognition~\cite{WeissZZSM21_DeepChromaChord_ISMIR}. 
Deep learning-based feature extractors yield the highest prediction accuracy when trained on data from the same distribution, which is, however, often not readily available. Thus, one
major challenge is the acquisition of a sufficient amount of correctly labeled training data. In classical music, it is often difficult to automatically annotate strongly aligned targets (short:~\emph{strong}~targets), i.e., with frame-wise target labels, due to changes of tempo. On the other hand, weakly aligned targets (short:~\emph{weak}~targets) only globally correspond to the input without containing frame-wise local alignments~\cite{WeissP21_DeepChromaMCTC_ISMIR,KrauseWM23_SoftDTW_ICASSP}. These weak targets are relatively easy to obtain, e.g., by only annotating start and end of an audio segment and deriving targets from the musical score. In our definition of weak targets, the order of the target vectors is correct, but their duration is unknown. Using weak targets in \ac{DNN} training requires a loss function that aligns network predictions with the corresponding weak targets.

In classification tasks, one widely used technique for training \acp{DNN} with weakly aligned targets is the \ac{CTC} loss~\cite{GravesFGS06_CTCLoss_ICML}, which aligns network predictions with a sequence of discrete labels. 
Despite being extendable to multi-label problems such as \ac{MPE}~\cite{WeissP21_MultiPitchMCTC_WASPAA}, \ac{CTC} remains limited to discrete targets and is algorithmically complex.

In contrast to \ac{CTC}, \ac{DTW} can be used to measure similarity between two real-valued sequences and has been successfully applied in, e.g., music synchronization and structure analysis~\cite{Mueller21_FMP_SPRINGER}. Recently, differentiable approximations of the minimum function~\cite{CuturiB17_SoftDTW_ICML,MenschB18_DifferentiableDynamicProgramming_ICML,HadjiDJ21_SmoothDTW_CVPR} have been included in \ac{DTW}, enabling the usage of the \ac{DTW} principle in gradient-based optimization algorithms. The algorithm proposed in~\cite{CuturiB17_SoftDTW_ICML}, \acf{SDTW}, uses so-called \emph{soft alignments} to compute a differentiable cost measure between sequences of different length. 
In~\cite{AgrawalWD21_ConvolutionalScoreAudioSync_SPL}, \ac{SDTW} is used in the context of performance-score synchronization and \cite{KrauseWM23_SoftDTW_ICASSP} employed \ac{SDTW} as a loss function to train \acp{DNN} for \ac{MPE} with weakly aligned pitch annotations. Experiments in~\cite{KrauseWM23_SoftDTW_ICASSP} indicated training instabilities with \ac{SDTW} when the sequence lengths of inputs and targets are significantly different. This poses a severe problem in many \ac{MIR} tasks, where sequences of input audio are typically very long, while weakly labeled targets, i.e., without note durations, are significantly shorter.

In this paper, we investigate the cause of training instabilities under the \ac{SDTW} loss and show that it is due to a mismatch between the estimated soft alignment and the reference alignment (see \cref{fig:teaser}a) in the early stages of training. This mismatch causes incorrect parameter updates and the training may diverge.
Therefore, we introduce and investigate strategies to decrease this alignment error to stabilize training. In particular, we analyze a hyperparameter scheduling strategy to yield smooth alignments in the early training phase (see \cref{fig:teaser}b) as well as the strategy of adding a diagonal prior to the \ac{SDTW} cost matrix to initially favour diagonal alignments (see \cref{fig:teaser}c). Furthermore, we investigate a sequence unfolding approach, where we uniformly stretch the weak target sequence to the length of the input sequence as proposed in~\cite{KrauseWM23_SoftDTW_ICASSP}. We choose \ac{DNN}-based \ac{PCE} as an exemplary task to study the training process of standard \ac{SDTW} and the impact of our stabilizing strategies. We demonstrate that the hyperparameter scheduling and the diagonal prior strategies reliably reduce label mismatch in the early training stage and therefore lead to successful trainings. In addition, these two strategies are computationally efficient and require only small modifications to the standard \ac{SDTW} algorithm. 

The remainder of this article is structured as follows. First, in \cref{sec:SDTW}, we discuss the \ac{SDTW} loss function and define the concept of soft alignments. Next, in \cref{sec:strategies}, we introduce three conceptually different strategies for stabilizing \ac{DNN} training under \ac{SDTW} loss. After describing the experimental setup in \cref{sec:expSetup}, we evaluate cause and effect of training problems with \ac{SDTW} in \cref{sec:eval}, along with the impact of our stabilizing strategies. Finally, we conclude with \cref{sec:conclusion} and give an outlook to potential areas of future research regarding \ac{SDTW}-based training in \ac{MIR}.

\section{Introduction to \acs{SDTW}}
\label{sec:SDTW}
In this section, we introduce \ac{SDTW} as a loss function in a \ac{DNN} training framework and define the concept of soft alignments, closely following~\cite{CuturiB17_SoftDTW_ICML,BlondelMV21_DiffDivergence_AISTATS}.

\subsection{Definition}
Let ${X=\left\{\mathbf{x}_0,\mathbf{x}_1,\dots,\mathbf{x}_{N-1}\right\}}$ denote a sequence of \ac{DNN} predictions, ${Y=\left\{\mathbf{y}_0,\mathbf{y}_1,\dots,\mathbf{y}_{M-1}\right\}}$ denote a sequence of weak targets and ${Y^\mathrm{S}=\left\{\mathbf{y}^\mathrm{S}_0,\mathbf{y}^\mathrm{S}_1,\dots,\mathbf{y}^\mathrm{S}_{N-1}\right\}}$ denote a sequence of strong targets, where $\mathbf{x}_n, \mathbf{y}_m, \mathbf{y}_n^\mathrm{S}\in\mathds{R}^D$ for $n\in\left\{0,1,\dots,N-1\right\}$ and $m\in\left\{0,1,\dots,M-1\right\}$. Without loss of generality, we assume $N\geq M$.

Using the \ac{MSE} as a local cost function, the elements of the cost \mbox{matrix}~${\mathbf{C}\coloneqq\mathbf{C}_{X,Y}\in\mathds{R}^{N\times M}}$ are computed as
\begin{equation}
\label{eq:costMatrix}
{\mathbf{C}_{X,Y}\left(n,m\right) = \lVert \mathbf{x}_n - \mathbf{y}_m\rVert_2^2}\,.
\end{equation}
We next define binary alignment matrices $\mathbf{A} \in\left\{0,1\right\}^{N\times M}$ which align two sequences of length $N$ and $M$. Each matrix $\mathbf{A}$ encodes an alignment via a path of ones from cell $(0,0)$ to $(N-1,M-1)$ using only vertical, horizontal, and diagonal unit steps~\cite{CuturiB17_SoftDTW_ICML}. All cells not corresponding to the alignment are set to zero. The set of all binary alignment matrices for sequences of length $N$ and $M$ is denoted $\mathcal{A}_{N,M}$.
Using a differentiable approximation of the minimum function
\begin{equation}
    \label{eq:softmin}
    \mathrm{softmin}^\gamma\left(\mathcal{S}\right) = -\gamma\log\sum_{s\in\mathcal{S}} \exp\left(-s/\gamma\right)
\end{equation}
for a given finite set $\mathcal{S}\subset \mathds{R}$ and a hyperparameter $\gamma\in\mathds{R}$, the \ac{SDTW} cost is given by
\begin{equation}
    \label{eq:sdtwCost}
    \mathrm{SDTW}^\gamma_{\mathbf{C}} = \mathrm{softmin}^\gamma\left(\left\{\langle \mathbf{A},\mathbf{C}\rangle,~\mathbf{A}\in\mathcal{A}_{N,M}\right\}\right)
\end{equation}
and can be computed efficiently via dynamic programming~\cite{CuturiB17_SoftDTW_ICML}. The inner product $\langle \mathbf{A},\mathbf{C}\rangle$ is the sum of all elements of $\mathbf{C}$ along the alignment given by $\mathbf{A}$. 
\subsection{Soft Alignments}
The expectation over all alignments $\mathbf{A}$ for a cost matrix $\mathbf{C}$ is captured by the soft alignment matrix~\cite{BlondelMV21_DiffDivergence_AISTATS}
\begin{equation}
\label{eq:softAlignmentMatrix}
    \mathbf{E}_{\mathbf{C}}^\gamma = \sum_{\mathbf{A}\in\mathcal{A}_{N,M}} p_{\mathbf{A},\mathbf{C}}^\gamma \mathbf{A} ~~~\in\mathds{R}^{N\times M}\,,
\end{equation}
where the probability of an alignment is defined as 
\begin{equation}
p_{\mathbf{A},\mathbf{C}}^\gamma = \frac{\exp\left(-\langle \mathbf{A}, \mathbf{C}\rangle/\gamma\right)}{\sum_{\mathbf{A^\prime}\in\mathcal{A}_{N,M}} \exp\left(-\langle \mathbf{A^\prime}, \mathbf{C}\rangle/\gamma\right)}\,.
\end{equation}
The soft alignment matrix is of particular interest as it is the the gradient of the \ac{SDTW} cost w.r.t.\ the local cost matrix
\begin{equation}
\label{eq:sdtwGradC}
    \nabla_{\mathbf{C}} \mathrm{SDTW}^\gamma_{\mathbf{C}} = \mathbf{E}_{\mathbf{C}}^\gamma
\end{equation}
and is computed during the backward pass of an \ac{SDTW} training step with a dynamic programming algorithm~\cite{CuturiB17_SoftDTW_ICML,BlondelMV21_DiffDivergence_AISTATS}. In contrast to the \emph{binary} alignments $\mathbf{A}$, the entries of the soft alignment matrix $\mathbf{E}_{\mathbf{C}}^\gamma \left(n,m\right)$ can be interpreted as the \emph{probability} of an alignment path going through cell $\left(n,m\right)$. 
Only if this soft alignment assigns probability mass to the correct alignments $(n,m)$, the local cost terms~\cref{eq:costMatrix} between the correct pairs of predictions $\mathbf{x}_n$ and targets $\mathbf{y}_m$ constitute the overall \ac{SDTW} cost and the \ac{DNN} parameters can be successfully trained.

The hyperparameter $\gamma$, also termed temperature, controls the smoothness of the softmin function \cref{eq:softmin}. Larger values of $\gamma$ lead to smooth minima in \cref{eq:sdtwCost}, i.e., with contributions of multiple alignments $\mathbf{A}$, and therefore a ``blurry'' soft alignment matrix $\mathbf{E}_{\mathbf{C}}^\gamma$ (see \cref{fig:teaser}b). On the other hand, small values of $\gamma$ promote ``sharp'' soft alignments $\mathbf{E}_{\mathbf{C}}^\gamma$ with fewer non-zero entries (see \cref{fig:teaser}a), as \cref{eq:softmin} converges to the hard minimum function in the limit $\gamma\rightarrow0$ and a single binary alignment $\mathbf{A}$ becomes dominant in \cref{eq:softAlignmentMatrix,eq:sdtwCost}.

\section{Stabilizing training with \acs{SDTW}}
\label{sec:strategies}
In this section, we introduce three strategies for stabilizing \ac{SDTW}-based training: hyperparameter scheduling, diagonal prior, and sequence unfolding.

\subsection{Hyperparameter Scheduling}
\label{sec:hypScheduling}
As described in \cref{sec:SDTW}, the softmin temperature parameter $\gamma$ controls the smoothness of the \ac{SDTW} soft alignments. While a low value of $\gamma$ is desirable to ensure exact correspondences between predictions and targets due to sharp alignments, the latter are problematic in the initial training phase as inaccurate predictions from randomly initialized network parameters lead to erroneous alignments, thus hampering convergence. Therefore, as a first strategy to stabilize \ac{SDTW} training, we discuss an epoch-dependent scheduling of $\gamma$. 
Starting a training with a large softmin temperature $\gamma^{\mathrm{start}}=10$ makes the soft alignment fuzzier, which leads to coarse, yet mostly meaningful target assignments (see \cref{fig:teaser}b). After ten epochs with $\gamma=10$, when the trained network predicts meaningful features, we linearly reduce $\gamma$ during the following ten epochs to a final value of $\gamma^{\mathrm{final}}=0.1$, which stays constant for the remaining training. 

\subsection{Diagonal Prior}
\label{sec:diagPrior}
\begin{figure}[t]
    \centering
    \begin{tikzpicture}
        	\begin{axis}[
        		width=.75\columnwidth,
        		height=.27\columnwidth,
        		scale only axis,
        		enlargelimits=false,
        		xtick={0,.2, .4, .6, .8, 1},
        		xmin=0,
        		xmax=1,
        		xticklabels={0, 100, 200, 300, 400, 500},
                ytick={0, .4, .8, 1.2, 1.6, 2},
                yticklabels={0, 10, 20, 30, 40,50},
        		ylabel={Target seq. (frames)},
                xlabel={Predicted sequence (frames)},
        		y label style={yshift=-.37cm, font=\footnotesize},
                x label style={yshift=0cm, font=\footnotesize},
                y tick label style={font=\footnotesize},
                x tick label style={font=\footnotesize},
        		clip=false,
                legend style={at={(1.0,1.1)}, anchor=south east, font=\footnotesize, legend columns=2},
                legend cell align=left,
                tick align=outside,
                ytick pos=left,
                xtick pos=bottom
        		]

                \addlegendimage{line legend, green, style={very thick}};
                \addlegendimage{line legend, red, style={very thick}};
          
        		\addplot graphics[xmin=0,xmax=1,ymin=0,ymax=2] {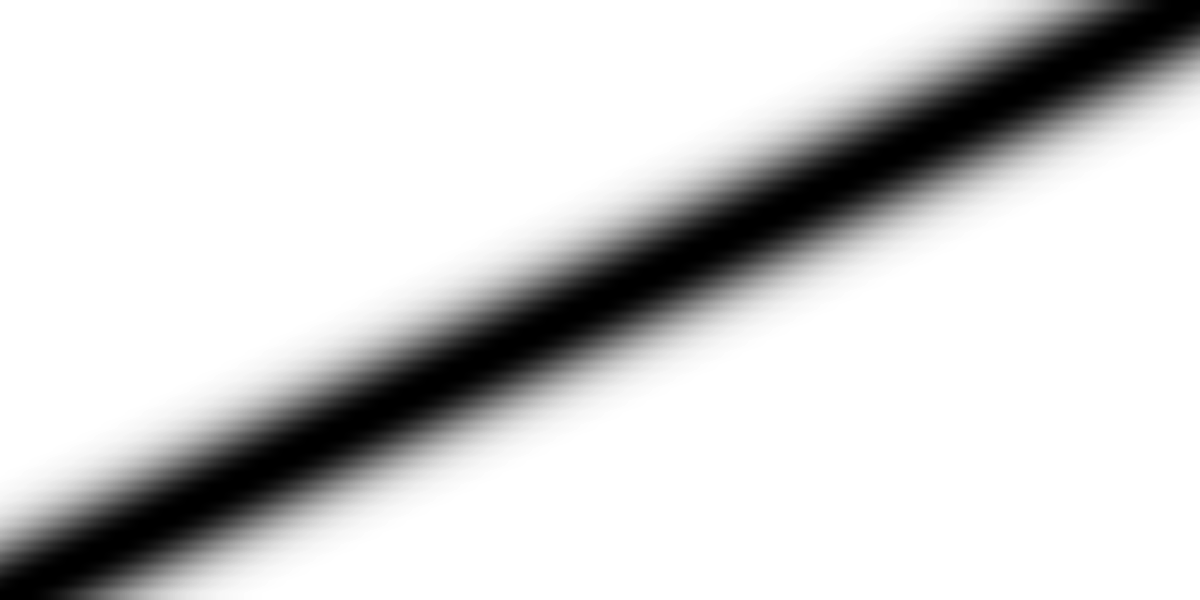};
        		\addplot graphics[xmin=1.05,xmax=1.1,ymin=0,ymax=2] {figs/cbar_bw_inv};
        		\node[anchor=west, font=\footnotesize] at (axis cs:1.1,0) {0};
        		\node[anchor=west, font=\footnotesize] at (axis cs:1.1,2) {1};
          
          \node[anchor=north, rotate=90, align=center, font=\footnotesize] at (axis cs:1.1,1.02) {Prior};



        	\end{axis}
        \end{tikzpicture}
    \vspace{-.7cm}
    \caption{Diagonal prior matrix $\mathbf{P}$ for $N=500$, $M=50$ and $\nu=1000$.}
    \label{fig:diag_prior_illustration}
    \vspace{-.3cm}
\end{figure}
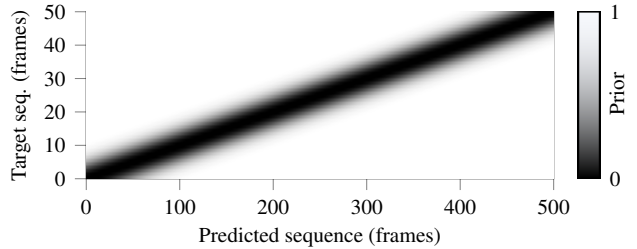

On average, the correct alignment of two sequences with arbitrary symbol durations has a higher probability to be close to the diagonal than to deviate from it. Therefore, as a second approach to stabilize the initial training phase, we investigate an additive prior ${\mathbf{P}\in\mathds{R}^{N\times M}}$ which penalizes elements of the cost matrix $\mathbf{C}$ that are far from the diagonal (see \cref{fig:diag_prior_illustration} for an illustration of a prior matrix). A similar strategy was employed in~\cite{ShihVBLPC21_radTTS_ICMLW} for restricting speech-text alignments to the diagonal.
Assuming equal symbol durations, the diagonal alignment of a target $\mathbf{y}_m$ starts at input frame $q_m=\lfloor \frac{Nm}{M}\rfloor$ and ends at $q_{m+1}-1$. To yield no penalty along the diagonal and a smoothly increasing penalty for distant alignments, we define the elements of the prior matrix as
\begin{equation}
    \mathbf{P}\left(n,m\right)=1-
    \begin{cases}
        1 ,& q_m\leq n < q_{m+1} \\
        \exp\left(\frac{\left(n-q_m\right)^2}{-2\nu}\right),& n < q_m\\
        \exp\left(\frac{\left(n-q_{m+1}\right)^2}{-2\nu}\right),& n \geq q_{m+1}\,,
    \end{cases}
\end{equation}
where the parameter $\nu$ controls the sharpness of the prior. In our experiments, we use $\nu=1000$. Finally, the prior matrix is added to the cost matrix with a weight $\omega$ to obtain the penalized cost matrix
\begin{equation}
    \mathbf{C}_{\mathbf{P}} \coloneqq \mathbf{C} + \omega \mathbf{P}\,,
\end{equation}
which replaces $\mathbf{C}$ in \crefrange{eq:sdtwCost}{eq:sdtwGradC}. Similarly to the hyperparameter scheduling strategy, we choose a constant prior weight $\omega=3$ during the first five epochs and then linearly reduce it to $\omega=0$ during the following five epochs.

Note that the numerical parameters for the strategies presented in \cref{sec:hypScheduling,sec:diagPrior} were determined empirically by the authors and small changes did not affect the training performance. However, when training on sequences of different length, with a different learning rate, or other \ac{DNN} types, parameters should be adjusted on a validation set. As presented in \cref{sec:eval}, analysis of the soft alignment matrix $\mathbf{E}_{\mathbf{C}}^\gamma$ provides a good indication of the current alignment stability.

\subsection{Sequence Unfolding}
Based on the observation that equal sequence lengths stabilize \ac{SDTW} training, a third strategy is to uniformly unfold the target sequence (see also~\cite{KrauseWM23_SoftDTW_ICASSP}). The unfolded target sequence ${Y^{\mathrm{U}}=\left\{\mathbf{y}_0^{\mathrm{U}},\mathbf{y}_1^{\mathrm{U}},\dots,\mathbf{y}_{N-1}^{\mathrm{U}}\right\}}$ is constructed by uniformly repeating elements from the weakly aligned target sequence, i.e., setting
\begin{equation}
    \label{eq:unfolding_elements}
    \mathbf{y}_n^{\mathrm{U}} \leftarrow \mathbf{y}_{\lfloor \frac{Mn}{N}\rfloor}
\end{equation}
to yield equal sequence lengths of the predictions $X$ and the targets $Y^{\mathrm{U}}$. Note that the repetition of target vectors introduces ambiguities, leading to multiple optimum alignments.

\section{Experimental Setup}
\label{sec:expSetup}
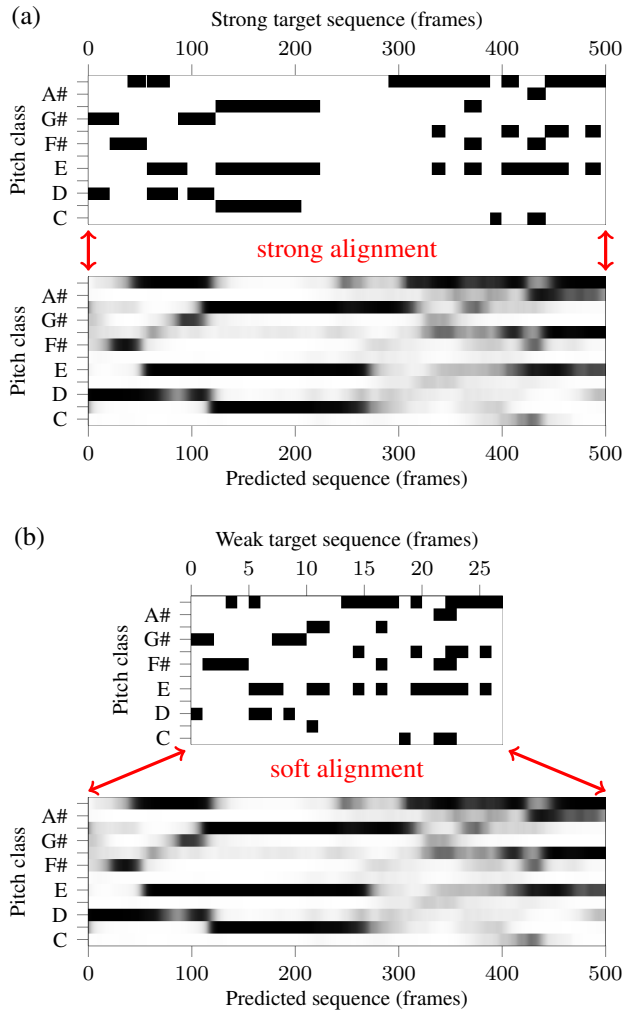
\begin{figure}[th!]
    \centering
    \begin{subfigure}[b]{\columnwidth}
        \centering
        \begin{tikzpicture}
            \begin{groupplot}[
            group style= {group size=1 by 2, vertical sep=.7cm},
            width=.83\columnwidth,
            height=.24\columnwidth,
              scale only axis,
              xmin=0,xmax=500,
              ymin=-0.5, ymax=11.5,
              ytick={0,1,2,3,4,5,6,7,8,9,10,11}, yticklabels={C,, D,, E,, F\#,, G\#,, A\#,},
              yticklabel style={font=\footnotesize},
             ylabel={Pitch class},
             y label style={yshift=-.3cm, font=\footnotesize},
             xticklabel style={font=\footnotesize},
             x label style={font=\footnotesize},
              tick align=outside,
              axis y line*=left,
              xlabel near ticks,
              clip=false
              ]
              \nextgroupplot[axis x line*=top,
              xlabel={Strong target sequence (frames)},
              x label style={font=\footnotesize, yshift=-.1cm},]
            \addplot graphics[xmin=0,xmax=500,ymin=-0.5,ymax=11.5] {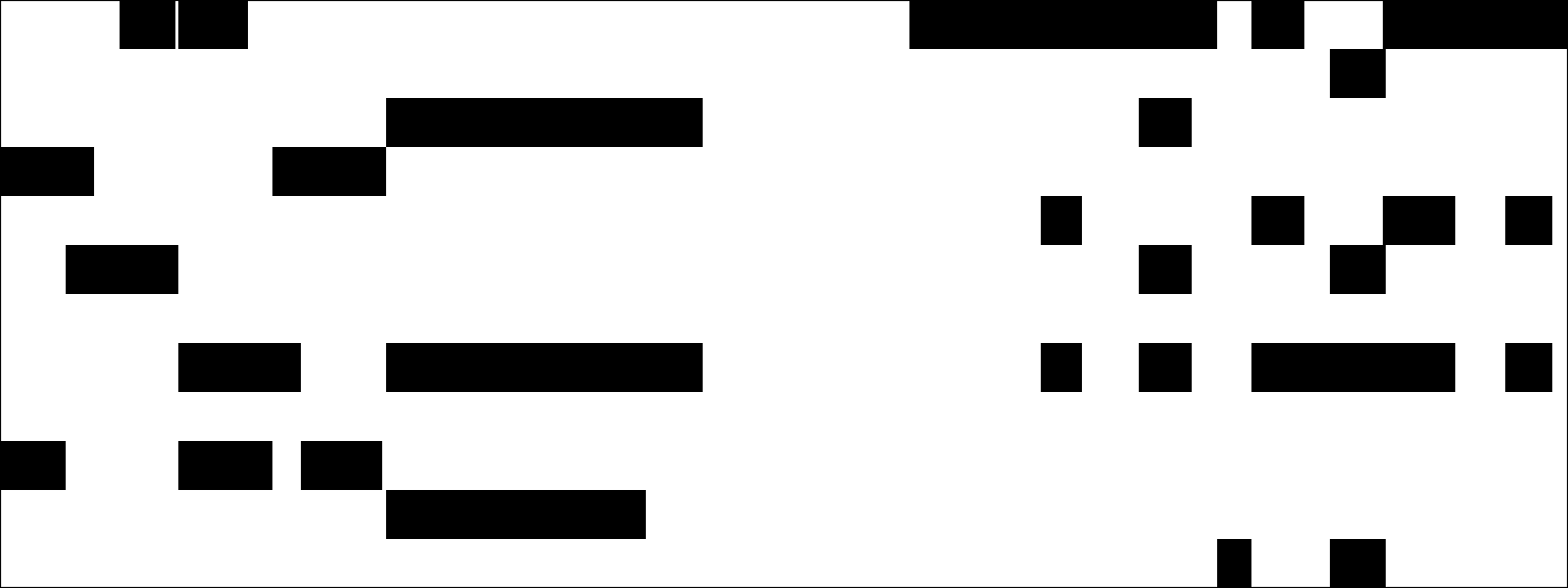};
            \node[anchor=south, align=right] at (axis cs: -60, 14.5) {(a)};
            \coordinate (top_leftBottom) at (axis cs:0, -1);
            \coordinate (top_rightBottom) at (axis cs:500, -1);
            \coordinate (top_centerBottom) at (axis cs:250, -1);

            \nextgroupplot[axis x line*=bottom,
            xlabel={Predicted sequence (frames)},
            x label style={font=\footnotesize, yshift=.1cm},]
            \addplot graphics[xmin=0,xmax=500,ymin=-0.5,ymax=11.5] {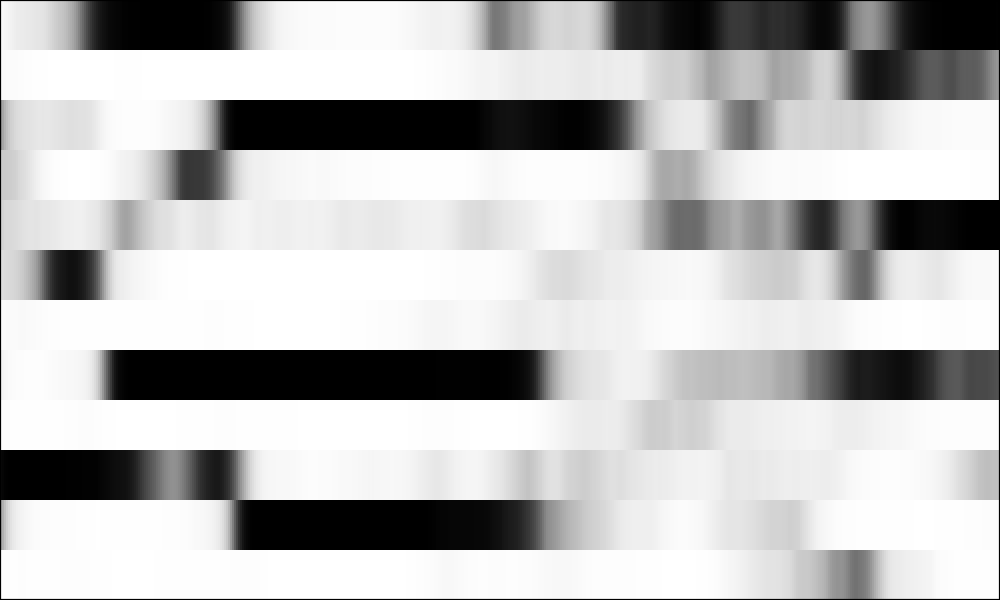};
            \coordinate (bottom_leftTop) at (axis cs:0, 12);
            \coordinate (bottom_rightTop) at (axis cs:500, 12);
            \coordinate (bottom_centerTop) at (axis cs:250, 12);
            
            \end{groupplot}     
            \draw[very thick, red, <->] (top_leftBottom) -- (bottom_leftTop);
            \draw[very thick, red, <->] (top_rightBottom) -- (bottom_rightTop);
            \node[red] (descr) at ($(bottom_centerTop)!0.5!(top_centerBottom)$) {strong alignment};
            
          \end{tikzpicture}
          \vspace{-.2cm}
    \end{subfigure}

    \begin{subfigure}[b]{\columnwidth}
        \centering
        \begin{tikzpicture}
            \begin{groupplot}[
            group style= {group size=1 by 2, vertical sep=.7cm},
            height=.24\columnwidth,
              scale only axis,              
              ymin=-0.5, ymax=11.5,
              ytick={0,1,2,3,4,5,6,7,8,9,10,11}, yticklabels={C,, D,, E,, F\#,, G\#,, A\#,},
              yticklabel style={font=\footnotesize},
             ylabel={Pitch class},
             y label style={yshift=-.3cm, font=\footnotesize},
             xticklabel style={font=\footnotesize},
             x label style={font=\footnotesize},
              tick align=outside,
              axis y line*=left,
              xlabel near ticks,
              clip=false
              ]
              \nextgroupplot[axis x line*=top,
              xmin=0,xmax=27,
              xtick={0,5,10,15,20,25},
              width=.5\columnwidth,
              xlabel={Weak target sequence (frames)},
              x label style={font=\footnotesize, yshift=-.1cm},]
            \addplot graphics[xmin=0,xmax=27,ymin=-0.5,ymax=11.5] {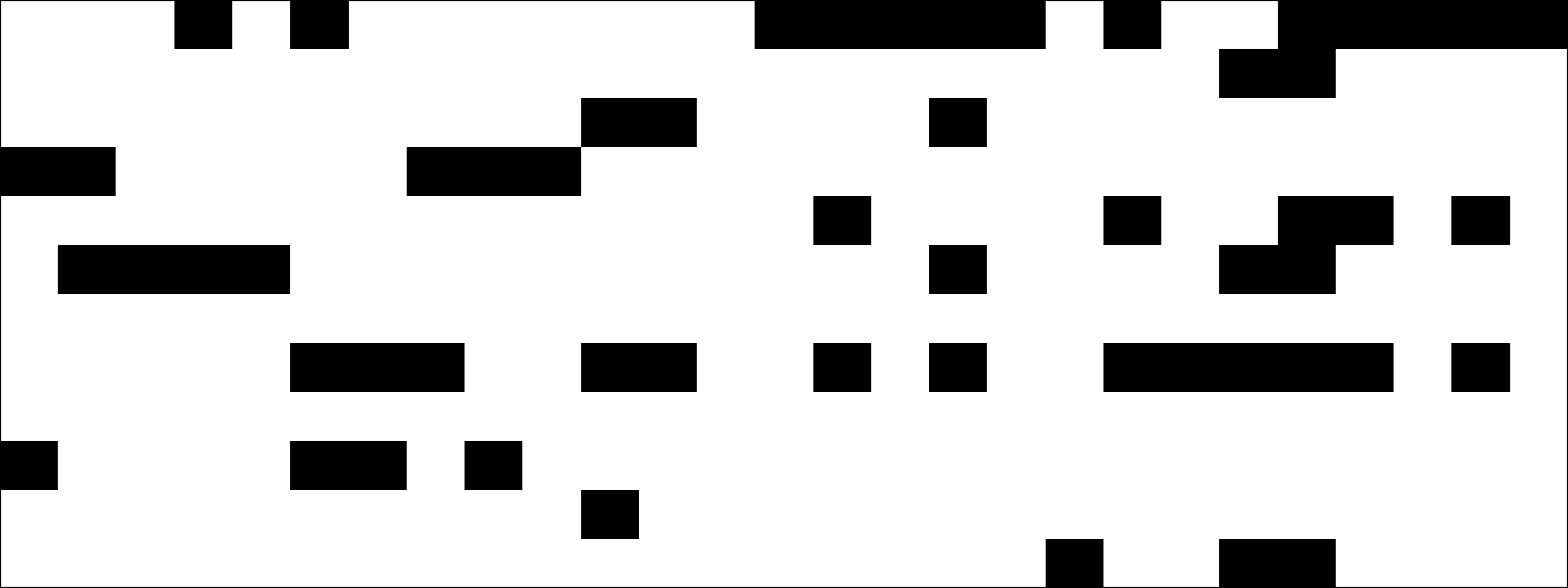};
            \node[anchor=south, align=right] at (axis cs: -14, 14.5) {(b)};
            \coordinate (top_leftBottom) at (axis cs:-.5, -1);
            \coordinate (top_rightBottom) at (axis cs:27.5, -1);
            \coordinate (top_centerBottom) at (axis cs:13.5, -1);

            \nextgroupplot[axis x line*=bottom,
            xmin=0,xmax=500,
            width=.83\columnwidth,
            xlabel={Predicted sequence (frames)},
            x label style={font=\footnotesize, yshift=.1cm},]
            \addplot graphics[xmin=0,xmax=500,ymin=-0.5,ymax=11.5] {figs/pitchClass_pred_runningEx};
            \coordinate (bottom_leftTop) at (axis cs:0, 12);
            \coordinate (bottom_rightTop) at (axis cs:500, 12);
            \coordinate (bottom_centerTop) at (axis cs:250, 12);
            
            \end{groupplot}     
            \draw[very thick, red, <->] (top_leftBottom) -- (bottom_leftTop);
            \draw[very thick, red, <->] (top_rightBottom) -- (bottom_rightTop);
            \node[red] (descr) at ($(bottom_centerTop)!0.5!(top_centerBottom)$) {soft alignment};
            
          \end{tikzpicture}
    \end{subfigure}
    \vspace{-.5cm}
    \caption{Alignment between training targets and predicted pitch class features $X$ for the running example from \emph{Frühlingstraum}. \textbf{(a)} Strong reference alignment for \acs{MSE} loss with strong targets $Y^\mathrm{S}$. \textbf{(b)} Soft alignment for \acs{SDTW} loss with weak targets $Y$.}
    \label{fig:PCE_illustration}
    \vspace{-.3cm}
\end{figure}

In this section, we describe the task for our case study, the employed dataset, as well as the used \ac{DNN} architecture and the training procedure.

\subsection{\acs{PCE} Task}
We choose \ac{PCE} from music recordings as an illustrative case study to investigate the problems of the \ac{SDTW} loss function and the effect of the stabilizing strategies. In our experimental setting, a \ac{DNN} takes $N$ frames of input audio (including context) and, for all frames, predicts twelve-dimensional pitch class activation vectors $X$ (see \cref{fig:PCE_illustration} for an illustration of predicted pitch class features). We want to train the \ac{DNN} such that the predictions $X$ match the training targets as close as possible. In the case of strong targets $Y^\mathrm{S}$, each predicted frame $\mathbf{x}_n$ is assigned to exactly one target frame $\mathbf{y}_n^\mathrm{S}$ using a strong alignment (see \cref{fig:PCE_illustration}a). When using weak targets $Y$, \ac{SDTW} internally computes a soft alignment based on the cost matrix $\mathbf{C}_{X,Y}$ to assign predictions and targets (see \cref{fig:PCE_illustration}b).

\subsection{Dataset}
Throughout all experiments, we use the \ac{SWD}~\cite{WeissZAMKVG21_WinterreiseDataset_ACM-JOCCH} which contains audio recordings and strongly aligned pitch class annotations. 
Winterreise is a song cycle for piano and singer, consisting of 24 songs. For each song, \ac{SWD} comprises nine different performances, resulting in $9\cdot24$ recorded songs with a total duration of \mbox{\SI{10}{\hour} \SI{50}{\minute}}.
We split the dataset for training, validation, and testing using a performance split~\cite{WeissZAMKVG21_WinterreiseDataset_ACM-JOCCH}. The publicly available performances by Huesch (HU33, recorded in 1933) and Scarlata (SC06, recorded in 2006) were annotated manually~\cite{WeissZAMKVG21_WinterreiseDataset_ACM-JOCCH} and constitute the test set. For training and evaluation we choose sequences of length $N=500$, corresponding to approximately $\SI{8.7}{\s}$ of audio at a sampling rate of $\SI{22050}{\hertz}$ and a hop length of $384$~samples. In order to generate weak training targets $Y$ from \ac{SWD} (which provides strongly aligned pitch class annotations $Y^\mathrm{S}$, see \cref{fig:PCE_illustration}a), we remove all adjacent repetitions of a pitch class vector (see \cref{fig:PCE_illustration}b)~\cite{WeissP21_DeepChromaMCTC_ISMIR}. We choose an excerpt from the song \emph{Frühlingstraum}, performed by Randall Scarlata (SC06), as a running example (see \cref{fig:PCE_illustration}) to visualize the soft alignment matrices (see \cref{fig:E_matrix_runningEx}).

\subsection{DNN Architecture and Training}
\begin{table}[t!]
    \small
    \centering
    \begin{tabular}{l l l l} 
         \hline
         Layer & Kernel Size & Stride & Output Shape \\
         \hline
         \multicolumn{4}{c}{\textbf{Prefiltering}} \\
         \hline
         LayerNorm & & & $(N+74,216,5)$ \\
         Conv2D & $15 \times 15$ & (1,1) & $(N+74,216,20)$ \\
         MaxPool & $3 \times 1$ & (1,1) & $(N+74,216,20)$ \\
         Dropout & & & \\
         \hline
         \multicolumn{4}{c}{\textbf{Binning to MIDI pitches}} \\
         \hline
         Conv2D & $3 \times 3$ & (1,3) & $(N+74, 72, 20)$ \\
         MaxPool & $13 \times 1$ & (1,1) & $(N+74, 72, 20)$ \\
         Dropout & & & \\
         \hline
         \multicolumn{4}{c}{\textbf{Time reduction}} \\
         \hline
         Conv2D & $75 \times 1$ & (1,1) & $(N, 72, 10)$ \\
         Dropout & & & \\
         \hline
         \multicolumn{4}{c}{\textbf{Chroma reduction}} \\
         \hline
         Conv2D & $1 \times 1$ & (1,1) & $(N, 72, 1)$  \\
         Dropout & & & \\
         Conv2D & $1 \times 61$ & (1,12) & $(N, 12, 1)$\\
         \hline
        \end{tabular}
    \caption{Musically motivated \acs{CNN} architecture~\cite{WeissZZSM21_DeepChromaChord_ISMIR,WeissP21_DeepChromaMCTC_ISMIR}.}
    \label{tab:cnn_architecture}
    \vspace{-.3cm}
\end{table}

We adapt a conceptually simple and musically motivated five-layer \ac{CNN} from\,\mbox{\cite{WeissZZSM21_DeepChromaChord_ISMIR,WeissP21_DeepChromaMCTC_ISMIR}} with 43383 trainable parameters to predict twelve-dimensional pitch class activation vectors from an input sequence. \Cref{tab:cnn_architecture} provides an overview of the architecture. 
We choose the \ac{HCQT}~\cite{BittnerMSLB17_DeepSalience_ISMIR} with five harmonics as an audio feature representation, spanning six octaves at a resolution of three bins per semitone (resulting in 216 frequency bins starting from C1), a hop length of $384$~samples and a frame rate of \SI{57.4}{\Hz}. From an input sequence of length $N+74$, the \ac{CNN} sequentially predicts $N$ vectors of pitch class activations. For the prediction of one frame, the \acs{CNN}'s receptive field covers 37 adjacent context frames on each side. 
Leaky ReLU with a negative slope of 0.3 is used as a nonlinearity after all hidden convolutional layers and sigmoid activation is used after the final layer. The dropout rate is set to 0.2.
All models are trained using the Adam optimizer~\cite{KingmaB15_OptimizerADAM_ICLR} with a batch size of 32 and an initial learning rate of 0.001.
We reduce the learning rate by a factor of two if the validation loss did not decrease during the last four epochs, and terminate the training if the validation loss did not decrease during the last twelve epochs. At the end of training, the model from the epoch with the lowest validation loss is restored. The source code for reproducing our experiments, as well as the trained models are available on \url{github.com/groupmm/stabilizing_sdtw}.
\section{Evaluation}
\begin{table}[t!]
\small
\begin{center}
\begin{tabular}{ c|c|c|c|c|c } 
\multicolumn{3}{c}{~}&&\multicolumn{2}{c}{F-measure} \\
 Loss&Targets & $\gamma$ & Strategy & mean & std \\
 \hline \hline
MSE&strong & - & - & \bf{0.82} & 0.07\\
 \hline \hline
 SDTW&weak & 0.1 & - & 0.56 & 0.37\\
  \hline
 SDTW&weak & 0.3 & - & 0.63 & 0.32\\
  \hline
 SDTW&weak & 1.0 & - & 0.24 & 0.36\\
  \hline
 SDTW&weak & 3.0 & - & 0.31 & 0.38\\ 
 \hline
 SDTW&weak & 10.0 & - & 0.57 & 0.37 \\
 \hline \hline
SDTW&weak &  $10 \rightarrow 0.1$ & {hyp. sched.} & 0.80 & 0.04 \\
 \hline
 SDTW&weak & 0.1 & diag. prior & \bf{0.81} & \bf{0.02} \\
 \hline
 SDTW&weak & 0.1 & seq. unfold. & 0.53 & 0.04 \\
 \end{tabular}
\caption{Averaged test results for \acp{DNN} trained on strongly aligned reference targets as well as \acp{DNN} trained with \ac{SDTW} on weakly aligned targets using either the standard configuration or the discussed stabilization strategies. We report the mean (higher is better) and standard deviation (lower is better) of the F-measure.}
\label{tab:resultsFM}
\end{center}
\vspace{-.5cm}
\end{table}
\label{sec:eval}
In this section, we investigate the training process as well as the prediction accuracy under the standard \ac{SDTW} loss, and compare it to the discussed stabilizing strategies. For quantitative evaluation, we repeat all \ac{DNN} trainings ten times from random initializations. For the test set predictions of each trained model, we compute the F-measure w.r.t.\ time-pitch class bins using a threshold of $0.5$. The mean and standard deviation of the F-measures from all trained models are displayed in \cref{tab:resultsFM}.
\subsection{Baseline: Strongly Aligned Targets}
As a first baseline and an upper bound for all following experiments, we consider \ac{DNN} training with strongly aligned targets $Y^\mathrm{S}$. For the sequence lengths ${M=N=500}$ and an \ac{MSE} loss function, the networks achieve the overall highest mean \mbox{F-measure} of $0.82$ with a standard deviation of $0.07$ on the test set.
\begin{figure}[t!]
     \centering
     \input{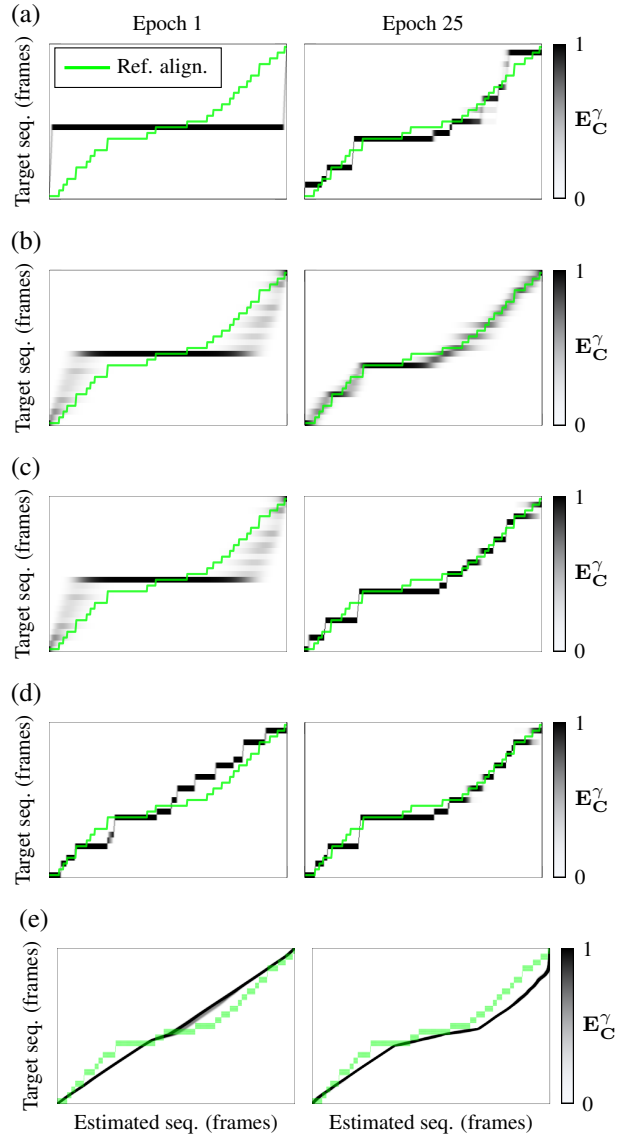}
        \caption{Reference alignment (green) and soft alignment matrix $\mathbf{E}_{\mathbf{C}}^\gamma$ (gray/black) for the running example after training epoch $1$ (left) and epoch $25$ (right) for different training strategies. \textbf{(a)} ${\gamma=0.1}$, \textbf{(b)} ${\gamma=10}$, \textbf{(c)} hyperparameter scheduling, \textbf{(d)} diagonal prior, \textbf{(e)} sequence unfolding.}
    \label{fig:E_matrix_runningEx}
    \vspace{-.6cm}
\end{figure}
\subsection{Standard \acs{SDTW}}
We next analyze \ac{DNN} training with weak targets $Y$ and the unmodified \ac{SDTW} formulation from~\cite{CuturiB17_SoftDTW_ICML,MaghoumiTL21_SoftDTWPytorch_IUI} as a loss function. We investigate five different values of \mbox{$\gamma\in\left\{0.1,\dots, 10\right\}$} which we keep constant during training. 
Analyzing the mean F-measure on the test set in \cref{tab:resultsFM}, the five variants with standard \ac{SDTW} yield comparably low results between $0.24$ and $0.57$, and high standard deviations between $0.32$ and $0.38$. Between $\SI{20}{\percent}$ ($\gamma=0.3$) and $\SI{70}{\percent}$ ($\gamma=1.0$) of all training runs converged to the all-zero output, indicating a highly unstable training process of standard \ac{SDTW}.
In order to determine the cause of these instabilities, we analyze the quality of automatically generated soft alignments in the \ac{SDTW} algorithm by visualizing the soft alignment matrix for the running example after training epochs one and 25, respectively. To highlight the effects of small and large values of $\gamma$, we focus on the edge cases $\gamma \in \left\{0.1, 10.0\right\}$. For $\gamma=0.1$, the estimated soft alignment exhibits a sharp structure (see \cref{fig:E_matrix_runningEx}a), which, after a collapse to a single target frame at epoch one, still only marginally overlaps with the reference alignment after 25 epochs. This sharp and erroneous soft alignment causes unstable gradient updates and leads to the collapse of many training runs. When choosing a large softmin temperature $\gamma=10$, \ac{SDTW} yields ``blurry'' soft alignments (see \cref{fig:E_matrix_runningEx}b) which at least partially capture the actual target frames in early epochs and coincide well with the reference alignments as training progresses. However, a blurry soft alignment also leads to blurry network predictions as multiple target frames are aligned to each predicted frame, thus resulting in a low F-measure when compared to strongly aligned targets.. 

\subsection{Stabilizing Strategies}
After evaluating the unsatisfactory training behavior of standard \ac{SDTW}, we investigate the effect of the previously introduced training strategies in the following section. We empirically choose $\gamma=0.1$ as the final softmin temperature in all following experiments, as sharp alignments are necessary for training an estimator with frame-wise precision.
\subsubsection{Hyperparameter Scheduling}
First, we combine the advantages of high and low values of $\gamma$ in a hyperparameter scheduling strategy. Starting a training with $\gamma=10$, the soft alignment matrix for our running example after one epoch is blurry and at least partially overlapping with the reference alignment (see \cref{fig:E_matrix_runningEx}c). The successive reduction to $\gamma=0.1$ until epoch 20 permits sharp alignments at a later training stage. Indeed, \cref{fig:E_matrix_runningEx}c shows a soft alignment after epoch 25 which is sharp and coincides well with the reference. The mean F-measure ($0.80$) in \cref{tab:resultsFM}, as well as the standard deviation ($0.04$), are the second best of all \ac{SDTW}-based trainings.
However, as the softmin function in \cref{eq:softmin} is a lower bound for the minimum function~\cite{HadjiDJ21_SmoothDTW_CVPR} which becomes tight for $\gamma\rightarrow 0$, the \ac{SDTW} loss is increasing when decreasing $\gamma$, despite unchanged network parameters. Therefore, this strategy does not allow for loss-based learning rate scheduling and early stopping before $\gamma$ is set to its final value. 
\subsubsection{Diagonal Prior}
\label{sec:experiments_diagPrior}
The second strategy stabilizes \ac{SDTW} trainings with low values of $\gamma$ by adding a penalty cost to \mbox{off-diagonal} \mbox{elements} of the cost matrix. For our running example in \cref{fig:E_matrix_runningEx}d, the soft alignment is indeed close to the diagonal after the first training epoch. As, on average, the alignments are diagonal, this often leads to correct assignments of predictions and targets even for randomly initialized \acp{DNN}. When the prior weight $\omega$ is reduced to zero after the initial training phase, the network is still able to adapt to off-diagonal alignments, as seen in our running example in \cref{fig:E_matrix_runningEx}d. Analyzing the performance metrics in \cref{tab:resultsFM}, using a diagonal prior yields the highest mean \mbox{F-measure} ($0.81$) and the lowest standard deviation ($0.02$) of all \ac{SDTW} variants, almost reaching the mean \mbox{F-measure} of the baseline experiments with strong targets and element-wise \ac{MSE} loss. Moreover, when the prior weight $\omega$ is reduced during training, the loss also decreases and therefore learning rate scheduling and early stopping are possible from the beginning.
\subsubsection{Sequence Unfolding}
Last, we investigate the strategy of unfolding the weak target sequence to the length of the input, which was employed in~\cite{KrauseWM23_SoftDTW_ICASSP}. For this strategy, we observe fully diagonal soft alignments in the initial training phase, as visualized for our running example in \cref{fig:E_matrix_runningEx}e. This is caused by the equal length of the predicted and the target sequence, which can be aligned using only diagonal steps. In the \ac{SDTW} formulation from~\cite{CuturiB17_SoftDTW_ICML}, the cost of a diagonal step is equal to the cost of a vertical or horizontal step. Thus, for a uniform cost matrix (which is probable at the initial training phase due to random network initialization), taking a diagonal step only accumulates half the cost compared to going ``around the corner'', i.e., one step in the vertical and one in the horizontal direction, or vice versa. This diagonalizing behavior leads, on average, to decent soft alignments in the early training phase (as discussed in \cref{sec:experiments_diagPrior}). However, in contrast to the additive diagonal prior strategy, the implicit diagonalization of alignments is not reduced during the training, as can be seen in \cref{fig:E_matrix_runningEx}e, which still exhibits strong diagonal components after 25 training epochs. Thus, the softly aligned \ac{SDTW} targets seldom match the reference targets and performance remains low, resulting in a mean F-measure of $0.53$ in \cref{tab:resultsFM}.

Note that the sequence unfolding strategy adds a significant computational overhead compared to the previous two strategies, as unfolding always corresponds to using a target sequence length of $M=N$. The forward and backward pass of the \ac{SDTW} loss function both have linear complexity w.r.t.\ the sequence lengths $\mathcal{O}\left(MN\right)$~\cite{CuturiB17_SoftDTW_ICML}. Thus, in our setting with $N=500$ and a mean length of the weak target sequences in the test set of $M=24$, the unfolding strategy leads to an increase in the computational cost of the \ac{SDTW} loss by a factor of more than 20.

\section{Conclusion and Outlook}
\label{sec:conclusion}
In this paper, we analyzed \ac{DNN} training instabilities with \ac{SDTW} as a loss function by the example of \ac{PCE}. By analysis of the soft alignment matrix, we argued that alignment mismatch in the early training phase often causes a collapse of the training procedure. Motivated by these findings, we investigated three strategies for stabilizing the early training phase. We found that the previously applied strategy of unfolding the weakly aligned target sequence leads to almost exclusively diagonal alignments due to a na\"{i}ve weighting of horizontal, vertical, and diagonal alignment steps. Furthermore, this strategy is computationally inefficient, as it increases the target sequence length. In contrast, the two introduced strategies of hyperparameter scheduling and diagonal prior can be implemented with negligible additional computational cost and stabilize \ac{SDTW}-based training by two different mechanisms. The hyperparameter scheduling strategy promotes smooth alignments in the early training phase, which increases the probability of the predicted frame being at least partially aligned to the correct target. Penalizing off-diagonal alignments in the \ac{SDTW} cost matrix by an additive diagonal prior is a strategy that initially restricts the soft alignment to a region of high probability. Experimental evaluation showed that these strategies reliably stabilize the \ac{SDTW} training process. Implementing them as a default in the \ac{SDTW} loss highly increases convergence rates.

Future research on \ac{SDTW}-based loss functions in \ac{MIR} applications might incorporate musically informed prior information, e.g., based on note durations or tempo annotations extracted from the musical score. Furthermore, the preference of diagonal alignment steps could be addressed by choosing different step weights.

\section{Acknowledgements}
This work was supported by the German Research Foundation (DFG MU 2686/7-2). The authors are with the International Audio Laboratories Erlangen, a joint institution of the Friedrich-Alexander-Universität Erlangen-Nürnberg (FAU) and Fraunhofer Institute for Integrated Circuits IIS.

\bibliography{referencesMusic,addBib}

\begin{thebibliography}{10}
\providecommand{\url}[1]{#1}
\csname url@samestyle\endcsname
\providecommand{\newblock}{\relax}
\providecommand{\bibinfo}[2]{#2}
\providecommand{\BIBentrySTDinterwordspacing}{\spaceskip=0pt\relax}
\providecommand{\BIBentryALTinterwordstretchfactor}{4}
\providecommand{\BIBentryALTinterwordspacing}{\spaceskip=\fontdimen2\font plus
\BIBentryALTinterwordstretchfactor\fontdimen3\font minus
  \fontdimen4\font\relax}
\providecommand{\BIBforeignlanguage}[2]{{%
\expandafter\ifx\csname l@#1\endcsname\relax
\typeout{** WARNING: IEEEtran.bst: No hyphenation pattern has been}%
\typeout{** loaded for the language `#1'. Using the pattern for}%
\typeout{** the default language instead.}%
\else
\language=\csname l@#1\endcsname
\fi
#2}}
\providecommand{\BIBdecl}{\relax}
\BIBdecl

\bibitem{BenetosDDE19_MusicTranscription_SPM}
E.~Benetos, S.~Dixon, Z.~Duan, and S.~Ewert, ``Automatic music transcription:
  {A}n overview,'' \emph{{IEEE} Signal Processing Magazine}, vol.~36, no.~1,
  pp. 20--30, 2019.

\bibitem{KorezeniowskiW16_DeepChroma_ISMIR}
F.~Korzeniowski and G.~Widmer, ``Feature learning for chord recognition: The
  deep chroma extractor,'' in \emph{Proceedings of the International Society
  for Music Information Retrieval Conference ({ISMIR})}, New York City, New
  York, USA, 2016, pp. 37--43.

\bibitem{WeissZZSM21_DeepChromaChord_ISMIR}
C.~Wei{\ss}, J.~Zeitler, T.~Zunner, F.~Schuberth, and M.~M{\"u}ller, ``Learning
  pitch-class representations from score--audio pairs of classical music,'' in
  \emph{Proceedings of the International Society for Music Information
  Retrieval Conference ({ISMIR})}, Online, 2021, pp. 746--753.

\bibitem{BartschW05_chroma_IEEEMULTIMEDIA}
M.~A. Bartsch and G.~H. Wakefield, ``Audio thumbnailing of popular music using
  chroma-based representations,'' \emph{{IEEE} Transactions on Multimedia},
  vol.~7, no.~1, pp. 96--104, 2005.

\bibitem{WeissP21_DeepChromaMCTC_ISMIR}
C.~Wei{\ss} and G.~Peeters, ``Training deep pitch-class representations with a
  multi-label {CTC} loss,'' in \emph{Proceedings of the International Society
  for Music Information Retrieval Conference ({ISMIR})}, Online, 2021, pp.
  754--761.

\bibitem{KrauseWM23_SoftDTW_ICASSP}
M.~Krause, C.~Wei{\ss}, and M.~M{\"u}ller, ``Soft dynamic time warping for
  multi-pitch estimation and beyond,'' in \emph{Proceedings of the {IEEE}
  International Conference on Acoustics, Speech and Signal Processing
  ({ICASSP})}, Rhodes Island, Greece, 2023.

\bibitem{GravesFGS06_CTCLoss_ICML}
A.~Graves, S.~Fern{\'{a}}ndez, F.~J. Gomez, and J.~Schmidhuber, ``Connectionist
  temporal classification: {L}abelling unsegmented sequence data with recurrent
  neural networks,'' in \emph{Proceedings of the International Conference on
  Machine Learning ({ICML})}, Pittsburgh, Pennsylvania, USA, 2006, pp.
  369--376.

\bibitem{WeissP21_MultiPitchMCTC_WASPAA}
C.~Wei{\ss} and G.~Peeters, ``Learning multi-pitch estimation from weakly
  aligned score-audio pairs using a multi-label {CTC} loss,'' in
  \emph{Proceedings of the {IEEE} Workshop on Applications of Signal Processing
  to Audio and Acoustics ({WASPAA})}, New Paltz, USA, 2021, pp. 121--125.

\bibitem{Mueller21_FMP_SPRINGER}
M.~M\"{u}ller, \emph{Fundamentals of Music Processing -- Using Python and
  Jupyter Notebooks}, 2nd~ed.\hskip 1em plus 0.5em minus 0.4em\relax Springer
  Verlag, 2021.

\bibitem{CuturiB17_SoftDTW_ICML}
M.~Cuturi and M.~Blondel, ``Soft-{DTW}: a differentiable loss function for
  time-series,'' in \emph{Proceedings of the International Conference on
  Machine Learning ({ICML})}, Sydney, NSW, Australia, 2017, pp. 894--903.

\bibitem{MenschB18_DifferentiableDynamicProgramming_ICML}
A.~Mensch and M.~Blondel, ``Differentiable dynamic programming for structured
  prediction and attention,'' in \emph{Proceedings of the International
  Conference on Machine Learning ({ICML})}, Stockholmsm{\"{a}}ssan, Stockholm,
  Sweden, 2018, pp. 3459--3468.

\bibitem{HadjiDJ21_SmoothDTW_CVPR}
I.~Hadji, K.~G. Derpanis, and A.~D. Jepson, ``Representation learning via
  global temporal alignment and cycle-consistency,'' in \emph{{IEEE/CVF}
  Conference on Computer Vision and Pattern Recognition ({CVPR})}, Virtual,
  2021, pp. 11\,068--11\,077.

\bibitem{AgrawalWD21_ConvolutionalScoreAudioSync_SPL}
R.~Agrawal, D.~Wolff, and S.~Dixon, ``A convolutional-attentional neural
  framework for structure-aware performance-score synchronization,''
  \emph{{IEEE} Signal Processing Letters}, vol.~29, pp. 344--348, 2021.

\bibitem{BlondelMV21_DiffDivergence_AISTATS}
M.~Blondel, A.~Mensch, and J.~Vert, ``Differentiable divergences between time
  series,'' in \emph{Proceedings of the International Conference on Artificial
  Intelligence and Statistics ({AISTATS})}, Virtual, 2021, pp. 3853--3861.

\bibitem{ShihVBLPC21_radTTS_ICMLW}
K.~Shih, R.~Valle, R.~Badlani, A.~{\L}a\'{n}cucki, W.~Piang, and B.~Catanzaro,
  ``{RAD-TTS}: Parallel flow-based {TTS} with robust alignment learning and
  diverse synthesis,'' in \emph{International Conference on Machine Learning
  ({ICML}), Third Workshop on Invertible Neural Networks, Normalizing Flows,
  and Explicit Likelihood Models}, Virtual, 2021.

\bibitem{WeissZAMKVG21_WinterreiseDataset_ACM-JOCCH}
C.~Wei{\ss}, F.~Zalkow, V.~Arifi-M{\"u}ller, M.~M{\"u}ller, H.~V. Koops,
  A.~Volk, and H.~Grohganz, ``{S}chubert {W}interreise dataset: {A} multimodal
  scenario for music analysis,'' \emph{{ACM} Journal on Computing and Cultural
  Heritage ({JOCCH})}, vol.~14, no.~2, pp. 25:1--18, 2021.

\bibitem{BittnerMSLB17_DeepSalience_ISMIR}
R.~M. Bittner, B.~McFee, J.~Salamon, P.~Li, and J.~P. Bello, ``Deep salience
  representations for {F0} tracking in polyphonic music,'' in \emph{Proceedings
  of the International Society for Music Information Retrieval Conference
  ({ISMIR})}, Suzhou, China, 2017, pp. 63--70.

\bibitem{KingmaB15_OptimizerADAM_ICLR}
D.~P. Kingma and J.~Ba, ``Adam: {A} method for stochastic optimization,'' in
  \emph{Proceedings of the International Conference for Learning
  Representations (ICLR)}, San Diego, California, USA, 2015.

\bibitem{MaghoumiTL21_SoftDTWPytorch_IUI}
M.~Maghoumi, E.~M. Taranta, and J.~LaViola, ``{DeepNAG}: Deep non-adversarial
  gesture generation,'' in \emph{Proceedings of the International Conference on
  Intelligent User Interfaces ({IUI})}, College Station, Texas, USA, 2021, pp.
  213--223.

\end{thebibliography}
 
\end{document}